\documentstyle[twocolumn,aps]{revtex}

\begin{document}
\draft
\title{Quantum Correlations in Two-Boson Wavefunctions}

\author{R. Pa\v skauskas and L. You}
\address{School of Physics, Georgia Institute of Technology,
Atlanta, GA 30332-0430}
\date{\today}
\maketitle

\begin{abstract}
We present the Schmidt decomposition for arbitrary
wavefunctions of two indistinguishable bosons, extending
the recent studies of {\it entanglement} or
{\it quantum correlations} for two fermion systems [J.
Schliemann {\it et al.}, Phys. Rev. B {\bf 63}, 085311 (2001)
and quant-ph/0012094]. We point out that the
von Neumann entropy of the reduced single particle
density matrix remains to be a good entanglement measure
for two identical particles.
\end{abstract}


Recently, quantum information and computing has emerged as
an active research field. Of the more fundamental issues in
quantum information science, characterization of inseparable
{\it quantum correlations}, or {\it entanglement} has received
considerable attention \cite{chuang}.
As is well accepted by now, the power of quantum computing
lies mostly at the inseparable nature of multi-partite
quantum states. In fact, it is often said that entanglement
is a quantum information processing resource (currency) which
all known quantum information protocols need to consume.
Historically, quantum entanglement
was first raised by Schr\"odinger \cite{sch} in responses
to EPR's local realism \cite{epr}. Despite more than 70
years of successful applications of quantum theory, however,
we still do not understand the full content of this
quantum inseparability.
At the simplest level of multi-partite pure states,
a complete knowledge for classifying and measuring
entanglement exists only if two distinguishable
particles are involved. For mixtures or states
of more than two particles, our current understanding
of inseparable correlations remains limited \cite{lewen}.

For quantum computing applications, generation of
entanglement is the prelude of any useful protocol \cite{chuang}.
One commonly adopted mechanism relies on a direct
interaction of the constituent quantum particles \cite{databus}.
For indistinguishable particles, e.g.
qubits represented by bosonic or fermionic atoms, photons, or electrons
as often envisioned in scalable quantum computer architecture \cite{qc},
a direct short-range interaction may cause significant overlap of
individual particle wavefunctions, thus compromising the ability of
keeping track of which particle is which qubit.
This calls for understanding inseparable quantum correlations
for identical particles including exchange symmetry.

The aim of this study is to investigate entanglement
measures for systems of two identical bosons. We extend the
recent results on two fermion systems by J. Schliemann
{\it et al.} \cite{tf1,tf2}, and present a unified
entanglement measure in terms of the von Neumann entropy
of the reduced single particle density matrix.
This paper is organized as follows: we first present
the well-known result for two distinguishable particles.
We then summarize the recent results of two fermions \cite{tf1,tf2}
and show that the von Neumann entropy of the reduced single particle
state remains a good entanglement measure.
Finally we provide a simple proof of the Schmidt decomposition
of two boson wavefunctions and obtain its von Neumann entropy
measure of entanglement in terms of the reduced single particle
density matrix. We conclude that in all three cases:
two distinguishable particles, two fermions or bosons,
the reduced single particle density matrix
fully characterizes their inseparable quantum correlations.

Physically, we arrive at the following conclusion:
for two identical particles, being simply in
usually inseparable quantum states does no necessarily
imply their use for quantum information science.
In agreement with the authors of Ref. \cite{tf2},
we will use {\it quantum correlations}, rather than
{\it entanglement}, to characterize useful
inseparable correlations of two identical particles
that are beyond what is normally
required by the permutation symmetry.

{\bf Two distinguishable particles}
For two distinguishable particles A and B,
any joint pure state wavefunction can be expressed as
\begin{eqnarray}
|\Psi\rangle=\sum_{i,j=1}^{N,M}C_{ij}a_i^\dag b_j^\dag |0\rangle,
\label{c}
\end{eqnarray}
where $C$ is a complex matrix ($N\!\times\!M$).
$a_i^\dag$ ($a_i$) and $b_j^\dag$ ($b_j$) are respectively
the single particle creation (annihilation) operators for A and B
in orthogonormal single states $|i\rangle_A$ (of dimension $N$)
and $|j\rangle_B$ (of dimension M). In the position
representation, one simply has $a_i^\dag|0\rangle\to \psi_i(\vec r_A)$
and $b_j^\dag|0\rangle\to \phi_j(\vec r_B)$.
The normalization condition gives ${\rm tr}(C^\dagger C)={\rm tr}(CC^\dagger)=1.$

The Schmidt decomposition of (\ref{c}) can now be constructed \cite{schmidt,can}.
Using the theorem of Singular Value Decomposition (SVD) for
a complex matrix \cite{book}, we write $C=U^TDV$ with unitary matrices
$U$ ($N\!\times\!N$) and $V$ ($M\!\times\!M$),
and a diagonal matrix $D$ [of dimension $\min(N,M)$].
Transforming the single particle basis according to
\begin{eqnarray}
a_k^{\prime\dag}&&=\sum_{i=1}^{N}U_{ik}^*a_i^\dag, \nonumber\\
b_{k}^{\prime\dag}&&=\sum_{j=1}^{M}V_{jk}^*b_j^\dag,\nonumber
\end{eqnarray}
the state vector (\ref{c}) becomes diagonal
\begin{eqnarray}
|\Psi\rangle=\sum_{k=1}^{\min(N,M)}D_{k}a_k^{\prime\dag}b_k^{\prime\dag}|0\rangle.
\end{eqnarray}
Therefore, it is said to be {\it entangled}
if more than one diagonal elements of $D$ ($D_k$) is nonzero,
i.e. if the rank of matrix $C$ is larger than 1.
A good entanglement measure in this case is the
von Neumann entropy of the reduced one particle density matrix.
The two particle density matrix is
$\rho=|\Psi\rangle\!\langle\Psi|$, and
the reduced density matrix is defined according to
$\rho^{A}\equiv {\rm tr}_B(\rho)$ and
$\rho^{B}\equiv {\rm tr}_A(\rho)$, whose elements can
be evaluated according to
\begin{eqnarray}
\rho^{A}_{\nu\mu}=\langle\Psi|a_\mu^\dag a_\nu|\Psi\rangle
=(C C^\dag)_{\mu\nu}^*,\nonumber    \\
\rho^{B}_{\nu\mu}=\langle\Psi|b_\mu^\dag b_\nu|\Psi\rangle
=(C^\dag C)_{\mu\nu}.
\end{eqnarray}
We immediately note that both \textit{Schmidt coefficients}
$|D_k|^2$ and the rank of matrix $C$ are independent
of unitary single particle transformations on $|\Psi\rangle$.
Furthermore, the von Neumann entropy
measure for entanglement
\begin{eqnarray}
S=&&-{\rm tr}_A[\rho_A\ln(\rho_A)]=-{\rm tr}_B[\rho_B\ln(\rho_B)]\nonumber\\
=&&-\sum_{k=1}^{\min(N,M)}|D_{k}^2|\ln|D_{k}^2|.
\end{eqnarray}
is also representation independent.
Because $0\leq|D_{k}^2|\leq 1$ and $\sum_k|D_{k}^2|=1$,
$S$ is zero for non-entangled states and takes a maximum
value $\ln[\min(N,M)]$ for the maximum entangled state.

{\bf Two-fermions}
A general two fermion state can be expressed as
\begin{eqnarray}
|\Psi_F\rangle=\sum_{i,j=1}^{N}\omega_{ij}f^{\dag}_if^{\dag}_j|0\rangle,
\label{f}
\end{eqnarray}
where $\omega$ is a complex and anti-symmetric matrix
($N\!\times\!N$) $\omega_{ij}=-\omega_{ji}$. $f_i^{\dag}$
($f_i$) is the creation (annihilation) operator of a fermion
in single particle state $|i\rangle$, and in position
representation $f_i^\dag f_j^\dag|0\rangle \to
[\psi_i(\vec r_A)\psi_j(\vec r_B)-\psi_i(\vec r_B)\psi_j(\vec r_A)]/\sqrt{2}$.
The normalization condition is
${\rm tr}(\omega^\dagger \omega)=1/2$.

As was first shown in Ref. \cite{tf1}, and extended in Ref.
\cite{tf2}, a Schmidt decomposition also exists in this case (\ref{f}).
A unitary transformation of $f_i=\sum_{j}U_{ij}f_j^{\prime}$ to
new fermionic operators $f_j^{\prime\dag}$ and $f_j^\prime$,
transforms $\omega$ in a non similar way.
The state (\ref{f}) becomes
\begin{eqnarray}
|\Psi_F\rangle=\sum_{i,j=1}^{N}\left(U^\dag\omega
U^\ast\right)_{ij} f^{\prime\dag}_if^{\prime\dag}_j|0\rangle.
\end{eqnarray}
There exists a matrix $U$ such that
$\omega$ takes a block diagonal form \cite{tf1,tf2,book},
containing $2\!\times\!2$ blocks of the type
\begin{eqnarray}
\left[
\begin{array}{cc} 0 & z_k \\ -z_k & 0\end{array} \right],
\end{eqnarray}
with eigenvalues $z_k$. Consequently the state (\ref{f})
becomes `diagonal' \cite{tf2}
\begin{eqnarray}
|\Psi_F\rangle=2\sum_{k=1}^{\le {N\over 2}}z_k
f^{\prime\dag}_{2k-1}f^{\prime\dag}_{2k}|0\rangle.
\end{eqnarray}
Several earlier studies on two electron correlations
in atomic systems also used this Slater-Schmidt
expansion \cite{kazik,kazik2}. In Ref. \cite{kazik},
Grobe {\it et al.} called it the `canonical representation'
(see also \cite{can}), and
a correlation measure was defined to be
$K=1/{\rm tr}(\rho^{f2})$. Approximately, $K$ measures the
number of Slater orbitals needed for the expansion of a
two electron wavefunction.
$K=1$, therefore, corresponds to no quantum correlation.

Schliemann {\it et al.} \cite{tf1,tf2} propose to use the
Slater rank, the number of non-zero $z_i$'s, as
the criterion for quantum correlations.
If the Slater rank is larger than 1, the two fermion state
$|\Psi_F\rangle$ contains additional quantum correlation
beyond the anti-symmetric permutation requirement.
Immediately, one sees that
there is no quantum correlations for two fermions in
two or three states $N=2,3$. The simplest case where additional
inseparable quantum correlations can exist occurs for $N=4$.
Several entanglement measures were proposed and studied
in Refs. \cite{tf1,tf2} for this case.

Compared to states for two distinguishable particles, we note
the analog of the von Neumann entropy remains a good correlation
measure for two fermions \cite{note}.
The two fermion density matrix is
$\rho_F=|\Psi_F\rangle\langle\Psi_F|$.
Because the two fermions are identical particles, there is
now only one single particle density matrix, $\rho^f$ (normalized to 1),
which can be computed according to
\begin{eqnarray}
\rho^f_{\nu\mu}&&={{\rm tr}(\rho_F  f_\mu^\dag f_\nu)\over
{\rm tr}(\rho_F\sum_\mu f_\mu^\dag f_\mu )}\nonumber\\
&&={1\over 2}\langle \Psi_F|f_\mu^\dag
f_\nu|\Psi_F\rangle=2(\omega^\dag\omega)_{\mu\nu}.
\end{eqnarray}
The normalization condition is thus reduced to simply
$\sum_{k=1}^{\le N/2}|z_k|^2=1/4$.
The corresponding von Neumann entropy is then
\begin{eqnarray}
S_f &&=-{\rm tr}[\rho^f\ln(\rho^f)]\nonumber\\
&&=-\ln2-4\sum_{k=1}^{\le {N\over 2}}|z_k|^2\ln(|z_k|^2).
\end{eqnarray}
It provides a smooth measure from the non-correlated state
$S_f=\ln 2$ to the maximum correlated state with $S_f=\ln(N_E)$.
$N_E$ is the largest even number not larger than $N$,
i.e. $N_E=N$ for even $N$ and $N_E=N-1$ for odd $N$.

We note that both the Slater rank and the von Neumann
entropy $S_f$ are independent of unitary
transformations on the single particle basis.

{\bf Two bosons}
We now study wavefunctions for two indistinguishable bosons
\begin{eqnarray}
|\Psi_B\rangle=\sum_{i,j=1}^{N}\beta_{ij}b^{\dag}_ib^{\dag}_j|0\rangle.
\end{eqnarray}
In this case $\beta$ a complex symmetric matrix ($N\!\times\!N$)
$\beta_{ij}=\beta_{ji}$.
$b_i^\dag$ ($b_i$) are the boson creation (annihilation)
operators for single particle state $|i\rangle$,
and in position representation $b_i^\dag b_{j\ne i}^\dag|0\rangle \to
[\psi_i(\vec r_A)\psi_j(\vec r_B)+\psi_i(\vec r_B)\psi_j(\vec r_A)]/\sqrt{2}$
and $b_i^\dag b_{i}^\dag|0\rangle \to \psi_i(\vec r_A)\psi_i(\vec r_B)$.
The normalization gives ${\rm tr}(\beta^\dag\beta)=1/2$.
Under a unitary transformation of $b$'s to new bosonic operators $b^\prime$'s
according to $b_i=\sum_jU_{ij}b_j^\prime$, the matrix $\beta$ undergoes
a non-similarity transformation $\beta=U^\dagger\beta U^{\ast}$. The
new state vector becomes
\begin{eqnarray}
|\psi_B\rangle=\sum_{i,j=1}^{N}\left(U^\dag\beta
U^\ast\right)_{ij} b^{\prime\dag}_ib^{\prime\dag}_j|0\rangle.
\label{lub}
\end{eqnarray}
We now prove that the Schmidt decomposition also exists in this
case. According to SVD \cite{book}, a complex symmetric matrix
needs only one unitary matrix $U$, such that $\beta=UBU^T$, and
$B$ being a diagonal matrix (with complex diagonal elements $B_k$).
The proof is rather simple. SDV gives
$\beta=UBV^T$, while $\beta$ being symmetric gives
$\beta=\beta^T=VBU^T$. Since SDV is unique, we obtain $U=V$
\cite{book2}.
In fact, the matrix U may be chosen to be the one that diagonalizes
$\beta$. Thus the state vector of two bosons
can be represented in a diagonal form
\begin{eqnarray}
|\Psi_B\rangle&&=\sqrt{2}\sum_{k=1}^{N}B_k b^{\prime\dag}_k b^{\prime\dag}_k|0\rangle.
\label{b}
\end{eqnarray}
We note that such a Schmidt expansion was suggested
earlier in Refs. \cite{kazik,kazik2}.
The inseparable correlations criterion is again related to
the number of non-zero diagonal elements $B_k$, i.e.
the rank of matrix $\beta$. In general quantum correlations
can arise as long as $N\ge 2$.

Again there is only one single particle density matrix
\begin{eqnarray}
\rho^b_{\nu\mu}={\langle\Psi_B|b^\dagger_{\mu}b_{\nu}|\Psi_B\rangle
\over  \langle\Psi_B|\sum_\mu b^\dagger_{\mu}b_{\mu}|\Psi_B\rangle}=2(\beta^\dagger\beta)_{\mu\nu}.
\end {eqnarray}
Similarly to the fermionic case, the
von Neumann entropy
\begin{eqnarray}
S_b=-{\rm tr}[\rho^b\ln\rho^b]=-\sum_{k=1}^N(2|B_k|^2)\ln(2|B_k|^2),
\end{eqnarray}
measures quantum correlations
between the non-correlated state $S_b=0$ and the maximum
correlated state $S_b=\ln(N)$.

Finally we note as discussed in Refs. \cite{tf1,tf2},
within a given rank, the determinant of the corresponding
non-zero sub-matrix can also be used to measure
quantum correlations. This was first shown in the simplest
case of $N=4$ for two fermions \cite{tf1} by introducing
a {\it dual} space (dual matrix for $\omega$). One can
show in general the sub-matrix determinant
takes the maximum value when $S_f$ or $S_b$ is the maximum.
We provide the simple case of two bosons below.
Let's assume the rank of $\beta$ matrix $r_b$ is equal to $N$.
We denote $|B_k|^2=\xi_k$ and construct a functional
${\cal{F}}=\prod_{k=1}^N\xi_k-\lambda(\sum_{k=1}^N\xi_k-1/2)$.
The second term is due to the requirement for a proper normalization
of the matrix $\beta^\dagger\beta$. We find the condition for
a maximum
by taking partial derivatives $\partial_{\xi_i}{\cal{F}}=0$.
The resulting equations then become
$\prod_{i\neq j}\xi_i=\lambda.$
Therefore the extreme occurs at
$\xi_1=\xi_2=\ldots=\xi_N={1}/{2N}.$ We may then express
the determinant of $\beta$ as
\begin{eqnarray}
{\cal F}=\xi_1(\xi_2,\xi_3,\ldots,\xi_N)\prod_{j=2}^{N}\xi_j.
\end{eqnarray}
Its second order derivative at $\xi_i={1}/{2N}$ is then
\begin{eqnarray}
\partial^2_{ij}{\cal{F}}=(\delta_{ij}-2)\left(\frac{1}{2N}\right)^{N-1},
\end{eqnarray}
which forms strictly negative matrix. Therefore the sub-matrix determinant
does take its maximum at this extreme point of $S_b$, a point which
corresponds to the maximally correlated state within the same rank.
Extending the proof to $r_b<N$ is straightforward.

Finally, we note that for two bosons our proposed entanglement measure
in terms of the single particle density matrix shares same spirit
with the so-called condensate fragmentation phenomena in
systems of many bosons \cite{james}. Therefore it may also provide
a useful starting point for the investigation of quantum correlations
beyond two particle systems. One obvious consequence for
many-boson systems is that fragmented condensates contain
non-zero inseparable quantum correlations or entanglement.

From a technical point of view, we note
Schmidt decompositions for two particle pure states
can be constructed starting from unitary transformations
that diagonalize the respective reduced single
particle density matrices.
We hope this unified approach will stimulate further discussions
into quantum information science of indistinguishable
particles. In the next paper, we will address the
question of such inseparable quantum correlations for
two bosons in quantum mixture states.

In conclusion, we have presented the Schmidt decomposition for
pure states of two indistinguishable bosons. We find the
rank of the Fock space wavefunction matrix can be used to
determine if additional quantum correlations (or entanglement)
beyond the usual permutation symmetry exists.
We have suggested the von Neumann entropy
for the reduced single particle density matrix as a smooth
entanglement measure for both two fermion and two boson wavefunctions,
just as in the case of two distinguishable particles.
There remains one point of concern in this von Neumann
entropy measure in that it did not seem to reflect
the information content of quantum correlations.
For instance, the un-correlated two fermion state gives
$S_f=\ln 2$, rather than 0 as in the classical case.
This points to the need for studies of quantum information
coding and processing with correlated identical particle states.

We thank YueHeng Lan and Dr. Peng Zhou for discussions.
This work is supported in part by the
NSF grant No. PHY-9722410 and the ARO/NSA grant G-41-Z05.

{\it Note added.}---After this paper was submitted for review,
a preprint \cite{lu} on the same topic of entanglement in a
two-identical-particle system appeared. It provided a
different proof of Eq. (\ref{lub}) in agalogy with the
original approach used for two fermions \cite{tf2}.
The reduced single particle density matrix
was not invoked in \cite{lu}. Although both results are
essentially the same in terms of mathematics,
we seem to arrive at different conclusions.
In particular, an interesting construction is provided in
Ref. \cite{lu} which transforms a symmetric state
$|B_k|(e^{i\phi_k}b_{k1}^{'\dag} b_{k1}^{'\dag}
+ e^{-i\phi_k}b_{k2}^{'\dag} b_{k2}^{'\dag})|0\rangle$
of a degenerate pair of non-zero diagonal element
$B_k=|B_k|e^{i\phi_k}$ into state
$b_{k1}^{\dag}b_{k2}^{\dag}|0\rangle$. Since both states
are of rank 2, we consider them as containing quantum
correlations. This is further evidenced by the fact that
both can be used for quantum teleportation \cite{tel}.
On the other hand, Ref. \cite{lu} considers them as
separable according to their definition 1.


\begin{references}
\bibitem{databus}This can be easily seen for a two partite system
with initial separable wavefunctions
$|\Psi(t=0)\rangle_{A+B}=|\psi\rangle_{A}|\phi\rangle_{B}$, a direct
interaction Hamiltonian $V_{AB}$ can lead to entanglement
(time evolution $\sim e^{-iV_{AB}t}$),
while separate unitary evolution from Hamiltonians $H_A$ or
$H_B$ does not.
\end{references}
\end{document}